# High Pixel Resolution Visible to Extended Shortwave Infrared Single Pixel Imaging with a black Phosphorus-Molybdenum disulfide (bP-MoS$_2$) photodiode

*Seyed Saleh Mousavi Khaleghi[1,2,3]\*, Jinyuan Chen[1]\*, Sivacarendran Balendhran[1], Alexander Corletto[1], Shifan Wang[1], Huan Liu[2,3,4], James Bullock[1], and Kenneth B. Crozier[1,3,4] ✉*

[1]Department of Electrical and Electronic Engineering, University of Melbourne, Victoria 3010, Australia

[2]University of Michigan-Shanghai Jiao Tong University Joint Institute, Shanghai Jiao Tong University, Shanghai 200240, China

[3]ARC Centre of Excellence for Transformative Meta-Optical Systems (TMOS), University of Melbourne, Victoria 3010, Australia

[4]School of Physics, University of Melbourne, Victoria 3010, Australia

\* Equal contribution

✉ e-mail: kenneth.crozier@unimelb.edu.au

ABSTRACT

High-resolution infrared imagers are currently more expensive than CMOS and CCD cameras, due to costly sensor arrays. Van der Waals (vdWs) materials present an opportunity for low-cost, room temperature infrared photodetectors. Although photodetectors based on vdWs materials show promising performance, creating a megapixel array is yet to be achieved. Imaging with a single vdWs photodetector typically relies on time-consuming mechanical scanning and suffers from low resolution. Single pixel imaging (SPI) offers an affordable alternative to achieve high-resolution imaging, utilizing only one photodetector and a spatial light modulator. Progress in SPI using vdWs material photodetectors has been limited, with only one prior demonstration in the near infrared range (64×64 pixels). In this work, we demonstrate a high-resolution SPI system (1023×768 for visible light and 512×512 for extended shortwave infrared) using a black phosphorus-molybdenum disulfide (bP-$MoS_2$) photodiode, surpassing earlier vdWs material SPI implementations by a factor of 64 in pixel count. We introduce an easy-to-implement edge detection method for rapid feature extraction. We employ compressed sampling and reduce imaging time by a factor of four. Our compressed sampling approach is based on a cyclic S-matrix, which is derived from a Hadamard-based sequence, where each row is a circular shift of the first row. This enables efficient imaging reconstruction via circular convolution and Fourier transforms, allowing fewer measurements while preserving the key image features. Our method for SPI using a vdWs material photodetector presents the opportunity for inexpensive shortwave infrared and midwave infrared cameras, and thus may enable advances in gas detection, biomedical imaging, autonomous driving, security, and surveillance.

Advanced imaging in the extended shortwave infrared (ex-SWIR: 1μm< λ < 3μm) and midwave infrared (MWIR: 3μm< λ < 8μm) is critical for applications such as biomedical diagnostics [1], gas detection [2], spectroscopy [3–5], autonomous systems [6], astronomy [7], security and surveillance [8]. Current technologies rely on costly photodetector arrays, and much interest exists in low-cost alternatives. Van der Waals (vdWs) material-based photodetectors such as those based on graphene, transition metal dichalcogenides and black phosphorus (bP) have shown much promise, offering broadband sensitivity, room temperature operation, and possible integration with CMOS processing [9–11]. While individual vdWs photodetectors rival commercial devices in detectivity and responsivity [12], [13], [14,15], scaling these into megapixel focal-plane arrays remains impractical due to material heterogeneity and integration challenges. Although there have been efforts to develop vdWs based focal plane arrays, no research has yet demonstrated a van der Waals-based megapixel focal plane array that spans the visible to MWIR range. Bianconi et al. [16] presented a promising solution-processed black phosphorus focal plane array with 0.07 megapixels. Wu et al. [17] demonstrated an 8×8 pixel array of $MoTe_2$, showcasing a facile thermal-assisted tellurization route for the growth of wafer-scale, phase-controlled 2D $MoTe_2$ layers. Xu et al. [18] introduced an 8×1 bP/$MoS_2$ vdWs photodetector linear array, fabricated using a temperature-assisted sloping transfer method. Single pixel imaging (SPI) circumvents this limitation by using a single photodetector paired with a spatial light modulator, enabling cost-effective imaging without array fabrication [19]. However, SPI systems leveraging vdWs material photodetectors remain underexplored, with prior works predominantly relying on mechanical scanning [20–26] or demonstrating imaging with modest pixel counts (64×64 pixel, [27]).

SPI reconstructs images by correlating the sampling patterns (e.g. as displayed on the system's spatial light modulator) with photodetector intensity measurements. Early SPI implementations stemmed from coded aperture principles [28,29], dual photography [30], ghost imaging [31], and compressed sensing [32–34]. Later implementations advanced to real-time video [35,36], multispectral imaging [37,38], and 3D imaging [39]. Despite progress, SPI faces challenges: low resolution (typically ≤256×256 pixels) [40,41], slow acquisition (limited by spatial light modulator refresh rate) [42,43], and noise limitations [44,45]. These constraints are amplified in infrared SPI, where expensive avalanche photodiodes or photomultipliers [46] are often required to compensate for the low signal-to-noise ratio.

Recent efforts to demonstrate SPI with vdWs material-based photodetectors have largely employed time-consuming mechanical scanning. For instance, Wang et al. [20] performed imaging of a transmission-type binary mask using a $WS_2$/Si photodetector by spatially scanning the image plane. Wang et al. [21], Wu et al. [22] and Zeng et al. [23] similarly scanned transmission masks with illumination wavelengths of 1.7 to 2.4 µm [21], 0.98 µm and 4.6 µm [22] and 4.55 µm and 10.6 µm [23], respectively. Tong et al. [24] extended this approach to polarized imaging, while Safaei et al. [25] extended this approach applied it to additional spectral regimes, including the longwave infrared. Chen et al. [47] also demonstrated scanning-based SPI with a high-mobility HgSe colloidal quantum dot photodetector, where a $BaF_2$ lens was mechanically scanned over a 15 × 15 mm² area to construct thermal images. The system sampled photocurrent at 3 kHz, achieving resolutions of 75 × 500 pixels, though the sequential nature of scanning inherently limits real-time performance. These previous works demonstrate the prevalence of scanning methods. Crucially, these works prioritize photodetector characterization over imaging innovation, and do not exploit the encoding advantages of SPI. In [26] Choi et al. demonstrated the use of a $MoS_2$ Schottky diode for single-pixel imaging, reporting that acquiring a 355 × 338-pixel image took approximately 17 hours due to the scanning method's stepwise measurement process (50 µm per step, averaged over 10 measurements). This underscores the efficiency limitations of scanning-based approaches for real-time applications, though the authors note that array-based sensor designs could enable faster acquisition. To date, only Miao et al. [27] attempted SPI with a two-dimensional material (a bP photoconductor) using a digital micromirror device (DMD) for pattern encoding. However, their system achieved a modest pixel count (64×64-pixel image) of a laser spot ($\lambda$= 0.83 µm), leaving unresolved challenges in signal-to-noise ratio and spatial resolution. Another challenge is that most vdWs material-based photodetectors have very small active areas [21,25,27], which are insufficient to capture the entire beam spot at the focal plane in an SPI setup.

High-resolution single-pixel imaging faces a challenge regarding acquisition time, as the combined duration of sampling and reconstruction scales with the image resolution. Minimizing this time is essential for practical applications. While compressed sensing techniques have been widely adopted to address this challenge, prior work has primarily focused on modest resolution systems (<256×256 pixels). Specifically, existing studies fail to bridge the gap between theoretical frameworks and real-world high-resolution implementation, where computational bottlenecks dominate. Several studies have provided foundational insights into compressive SPI frameworks. For example, Gibson et al. [48] reviewed early development in SPI, while Zhao et al. [49], Lu et al. [50] and Bian et al. [51] examined various classes of measurement matrices (e.g., random, semi-deterministic, and partial orthogonal) and reconstruction algorithms (e.g., convex optimization, greedy methods, and total variation minimization). However, these studies do not address the inefficiency of optimization-based reconstruction at high resolutions, where iterative methods become prohibitively slow due to computational complexity. This gap is exacerbated when using high-speed spatial light modulators (e.g., DMDs): the time spent on reconstruction often exceeds the measurement time itself, nullifying hardware advantages. While deep learning methods [35,44,52] offer faster alternatives, prior work has not sufficiently explored their integration with high-resolution SPI systems, nor has it provided scalable solutions for balancing speed and reconstruction quality.

Edge detection is crucial for machine vision, autonomous systems, and biomedical imaging, with SPI emerging as a promising alternative to conventional pixel-array sensors. Previous SPI approaches include gradient ghost imaging (GGI) [53] which requires prior knowledge on the object, speckle-shifting methods [54] that avoid the need for this prior knowledge, and techniques using structured illumination [55], frequency-domain processing [56], spiral-phase encoding [57], or modified Fourier patterns (Laplacian of Gaussian operators) [58]. All the aforementioned methods involve complex sampling pattern optimization or reconstruction that hinder practical implementation.

Here, we demonstrate high resolution SPI using a bP-MoS$_2$ heterostructure photodiode, addressing key limitations of prior vdWs material photodetector-based SPI systems. Our contribution is fourfold. 1. High resolution imaging: Our system achieves unprecedented resolution for visible (1023×768 pixels) and extended short wave infrared (512×512 pixels) SPI. More than 64× improvement in pixel count over prior 2D-material-based SPI demonstrations (typically <64×64 pixels [27]). This leap in resolution is enabled by making a large active area photodetector and a scalable optical design that addresses the size mismatch problem that limits conventional vdWs detectors. 2. Easy to implement and rapid edge detection: Unlike conventional edge detection methods that require complex sampling patterns or prior knowledge of the target object, we demonstrate a straightforward electrical solution. By applying a high-pass filter through a low-noise current preamplifier to the bP-MoS$_2$ photodiode output, we directly extract image edges without computational reconstruction or pattern optimization. This approach eliminates the implementation barriers of existing techniques while maintaining real-time operational capability.

3. Cyclic S-matrix compressed sampling technique: To address the challenges of high-resolution single-pixel imaging identified above, particularly the computational bottlenecks and inefficiencies of traditional optimization-based reconstruction, we developed a cyclic S-matrix compressed sampling technique tailored for full-resolution DMD systems. Unlike prior methods that become prohibitively slow at high resolutions, our approach significantly reduces acquisition time by a factor of four, while maintaining acceptable reconstruction quality, with peak signal-to-noise ratio (PSNR) above 20 dB and structural similarity index measure (SSIM) above 0.22 relative to the non-compressed image. This technique effectively leverages the high-speed modulation capability of DMDs by ensuring that reconstruction time does not outweigh measurement speed, thus overcoming a key limitation in existing systems. 4. Spectral sensitivity: To show an application of the fact that our system extends to the shortwave infrared, we demonstrate imaging through an object opaque to visible light, namely a silicon wafer. Although visible/SWIR SPI through silicon has been achieved with conventional photodetectors [59], our work establishes the first demonstration of SPI through silicon using a vdWs material-based photodetector. This capability could be useful for non-destructive testing and machine vision. Our work establishes vdWs material photodetectors as viable candidates for SPI, bridging the gap between novel device development and functional imaging systems.

**Imaging Setup and result:**

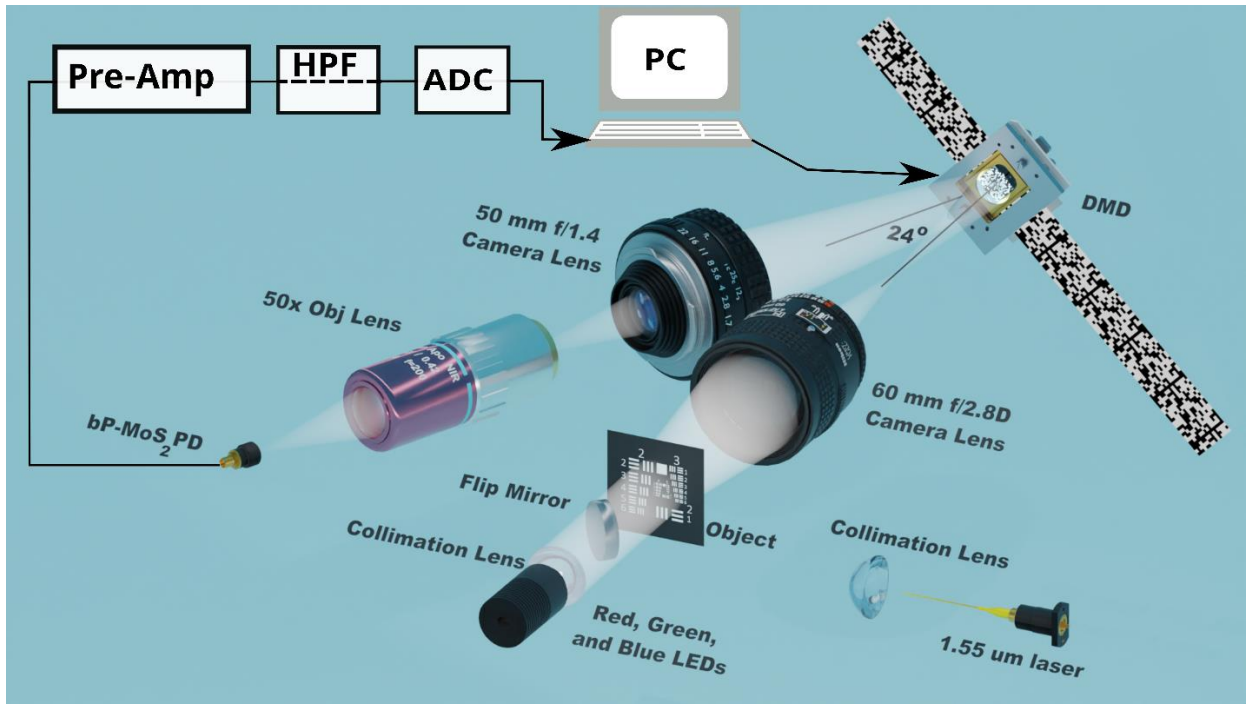

Figure 1. 3D schematic of SPI setup based on vdWs material photodetector. Abbreviations used here: PD: photodiode, DMD: digital micromirror device, Obj: objective lens, Pre-Amp: low noise current pre-amplifier, HPF: high pass filter, ADC: analog to digital converter, PC: personal computer.

A schematic of the developed SPI system is provided as Figure 1. It can be considered to comprise four submodules, namely for imaging, modulation, detection, and processing. The imaging module uses visible-wavelength LEDs ($\lambda$ =0.45 μm, 0.565 μm, 0.78 μm) and an infrared laser ($\lambda$ =1.55 μm), collimated to illuminate test objects in transmission mode. Test objects include a glass slide with gold film patterned with the image of a kangaroo, color photographic slides (35 mm), and a USAF 1951 resolution target (chrome on glass). The transmitted light is imaged by a camera lens (Nikon, focal length 60 mm) onto the DMD at normal incidence. The modulation module consists of the DMD (V-7000, Vialux). As we discussed later, we configure this to display cyclic S-matrix patterns by scrolling, which reduces memory load and acquisition time. The detection module employs a two-stage demagnification system, comprising a camera lens (Pentax, focal length: 50 mm) to collect light reflected by the DMD and a microscope objective lens (Mitutoyo, 50× magnification). The two lenses are positioned at an optimized separation distance to generate a tightly focused spot after the 50× objective lens. This is explained in the Supplementary Information, where Section I approximately models the illuminated area at the focal plane and the effect of the small active area of the photodetector on the reconstructed image. The photodiode is connected to a low-noise current preamplifier with zero bias applied. The preamplifier measures the short-circuit current and converts it to an output voltage. The output voltage of the preamplifier is digitized by a 14-bit analog-to-digital converter (ADC). A personal computer serves as the processing module, handling system coordination. It generates and loads cyclic S-matrix patterns to the DMD, synchronizes ADC sampling with DMD triggers, and reconstructs images based on photocurrent-pattern correlations.

Achieving high resolution with vdWs material photodetectors faces three main constraints. 1. In prior studies, vdWs photodetectors have typically featured limited active areas (< 30μm ×30μm), which were smaller than the beam spot at the focal plane [21,24,25,60,61]. When used in SPI, this mismatch leads to incomplete signal capture and a loss of high-spatial-frequency information. 2. Insufficient photodetector bandwidth, which increases per-sample acquisition time, leading to longer total acquisition times in high-resolution imaging [20,22]. 3. Noise equivalent power (NEP) values that are high, leading to reconstructed images of poor quality.

To overcome the challenge of limited active area in vdWs material-based photodetectors, we fabricated a large-area bP-MoS$_2$ photodiode and assessed its suitability for SPI. Specifically, we conducted responsivity measurements, noise current density measurements, and photocurrent response tests under different DMD switching speeds. Figure 2 (a) shows the device mounted on a 28-pin chip carrier, including a zoomed-in optical image. Figure 2 (b) illustrates the architecture of the bP-MoS$_2$ photodiode. Section II of the Supplementary Information explains the details of the fabrication process. We measured the responsivity of the device under zero bias at room temperature in air across the visible to midwave infrared range using two different methods, with the results shown in Figure 2(c). For the visible to SWIR range, we illuminated the zero-biased bP-MoS$_2$ photodiode with light from a monochromator and measured the resulting photocurrent. To determine the incident light intensity at the detector plane, we calibrated the optical system using reference Si and Ge photodiodes with known responsivities through a 100 µm-diameter pinhole. The measured photocurrent from the bP-MoS$_2$ device, along with the calibrated light intensity, was then used to calculate its responsivity. Results are shown in the left panel of Figure 2(c). For the ex-SWIR to MWIR range, we measured the photocurrent of bP-MoS$_2$ photodiode using a Fourier transform infrared (FTIR) spectrometer. To determine its spectral responsivity, we employed a calibrated measurement protocol. First, we recorded the FTIR signal using a deuterated triglycine sulfate (DTGS) photodetector, which has a flat spectral response across the range of interest. We then replaced the DTGS detector with the bP-MoS$_2$ photodiode and measured its photocurrent under identical illumination conditions and zero bias. By taking the ratio of the bP-MoS$_2$ signal to that of the DTGS detector at each wavelength, we obtained the device's relative spectral responsivity. For absolute calibration, we repeated the measurements using a reference InGaAs photodiode with a known responsivity spectrum. By scaling the relative responsivity of the bP-MoS$_2$ photodiode using the InGaAs-to-DTGS responsivity ratio, we derived its absolute spectral responsivity. The results are shown in the right box of Figure 2(c). The bP/MoS$_2$ device exhibits a maximum responsivity of 0.066 A/W at λ = 0.68 μm, with sharp declines across MoS$_2$'s absorption range (λ = 0.68 – 0.78 μm), where absorption occurs in both bP and MoS$_2$ layers, and beyond λ = 4 μm, the absorption limit of bP. For λ > 0.78 μm, the MoS$_2$ layer becomes transparent, and absorption occurs only in bP layer. No response was detected beyond 4.4 μm, consistent with prior studies on bP-MoS$_2$ photodiodes [12,21,60]. Unlike earlier work, our measurements extend to the visible range, offering comparative insights. The 30× lower responsivity in SWIR/MWIR compared to visible spectrum necessitates 30× higher light intensity for single-pixel imaging in SWIR/MWIR range. In addition, we characterized the noise current density, defined as the root-mean-square noise current per square root bandwidth ($A/\sqrt{Hz}$), of the bP-MoS$_2$ photodiode connected with the preamplifier in SPI setup Figure 2 (d). The measurements were conducted at the maximum operating frequency of the DMD (22.727 kHz). The results show that beyond this frequency, the noise current density remains below $2\ pA/\sqrt{Hz}$. Based on the measured current-voltage (I-V) characteristics, the resistance of the photodiode at zero bias is about 13.191 MΩ. The Johnson noise for the photodiode is predicted to be $0.03\ pA/\sqrt{Hz}$. Notably, our bP-MoS$_2$ photodiode remains functional after 1.5 years, with most of the data collected post-1-year fabrication. All the measurement reported here has been done in ambient condition, demonstrating superior stability of the device.

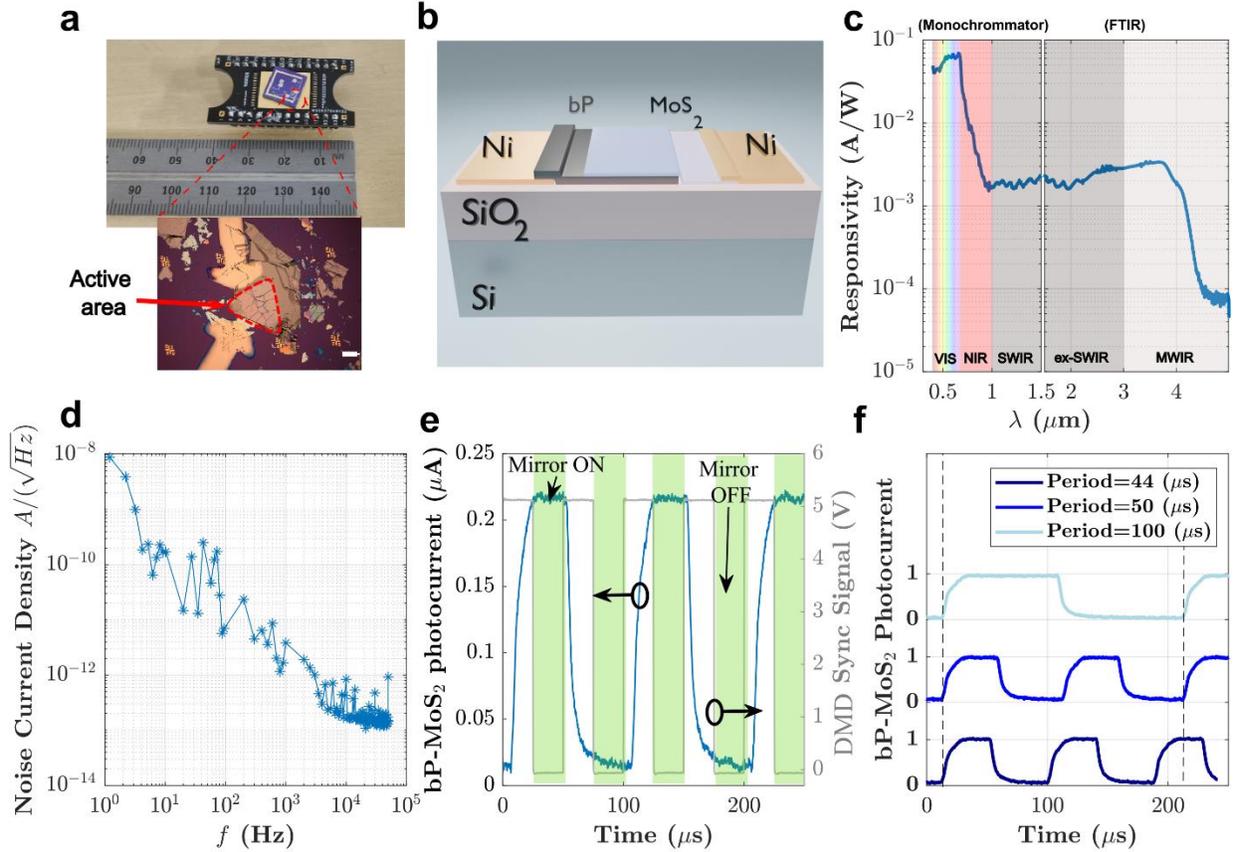

Figure 2. (a) Photograph of bP-MoS$_2$ photodiode mounted on standard 28-pin chip carrier (top panel). Optical microscope image of photodiode (lower panel). White scale bar is 80 µm. (b) Schematic illustration of bP-MoS$_2$ photodiode. (c) Responsivity of bP-MoS$_2$ photodiode, measured using monochromator (left panel) and FTIR (right panel). (d) Noise current density of bP-MoS$_2$ photodiode connected to low noise current preamplifier, measured by lock-in amplifier. (e) Photocurrent vs time (left y-axis), for on-off key modulation by DMD. DMD synchronization signal vs time (right y-axis). Green boxes indicate the intervals during which data is acquired (stable mirror state), thereby excluding transition periods when mirrors are in motion. (f) Normalized photocurrent of bP-MoS$_2$ photodiode versus time at different switching periods of DMD.

To synchronize the DMD switching pattern with the ADC measurement signal, we use the DMD synchronization signal. The latter is shown along with the photocurrent in Figure 3 (e). These are measured during an experiment with on/off key modulation produced by driving the DMD with black and white sample patterns at a frequency of 20 kHz. The white patterns comprise all mirrors in the "on" position, i.e. reflecting light to the photodetector. The black patterns comprise all mirrors in the "off" position, directing light away from the photodetector. The bP-MoS$_2$ photocurrent is shown as the blue curve, while the gray curve shows the DMD sync signal. Data sampling occurs exclusively during the green-shaded intervals, which corresponds to periods of stable micromirror alignment. To verify that the photodiode's 3 dB bandwidth exceeds the DMD's maximum modulation frequency (22.727 kHz), we measured photocurrent response at switching periods of 100 μs, 50 μs, and 44 μs, Figure 2. (f). The signal reached steady-state with consistent amplitude across all tested periods, confirming that the large-area photodiode maintains sufficient bandwidth (>22.727 kHz). This result agrees with previously reported bandwidth measurements of bP-MoS$_2$ photodiodes [12].

Having characterized the responsivity, noise performance and speed of the bP-MoS$_2$ photodiode, we next demonstrate its imaging capabilities across a broad spectral range. To do so, we captured images spanning from the visible to the ex-SWIR range using our SPI setup. Figure 3 shows images captured in the visible and ex-SWIR ranges at maximum light intensities across various wavelengths. By capturing images of the source using both the bP-MoS$_2$ photodetector and a commercial photodiode in the visible and ex-SWIR ranges, we observe that the commercial detector produces uniformly white images across the entire field in both spectral ranges. Similarly, the bP-MoS$_2$ detector yields a uniformly white image in the visible range. However, in the ex-SWIR image acquired with the bP-MoS$_2$ device, only a central portion corresponding to approximately 512×512 pixels on the DMD appears white, while the surrounding regions remain black. Figure 3(a) displays a kangaroo image captured with the laser operating at λ =1.55 µm. There are some black dots in the kangaroo body which were caused by some residue on the glass slide.

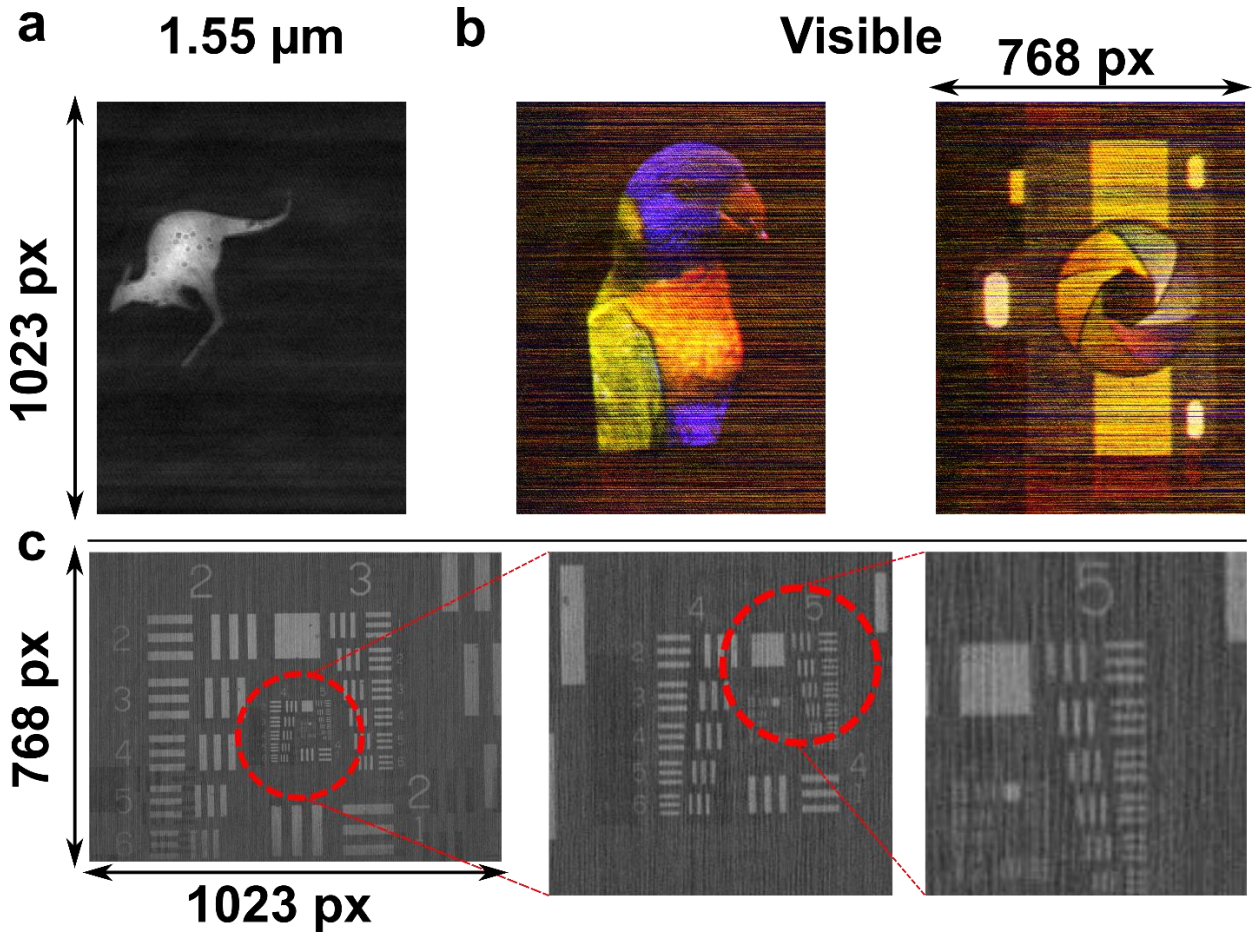

Figure 3. (a) Image of glass slide with gold film patterned with a kangaroo silhouette captured via 1550nm laser. (b) Images of photographic slide (35 mm film) of bird (lorikeet) and camera icon, captured under sequential illumination by red, green, and blue LEDs. (c). Image of USAF1951 resolution target. Right panels show zoom-ins of the resolution target.

To demonstrate visible-wavelength SPI with our bP-MoS$_2$ photodiode, we used an object comprising a color photographic slide (35 mm). One slide contained the image of an Australian bird (lorikeet), and another contained an image of a camera (Figure 3(b)). To generate RGB images, we employed multispectral imaging by capturing separate exposures under narrowband LED illumination ($\lambda$ =0.45 μm, 0.565 μm, 0.78 μm). Since these wavelengths do not align with standard RGB primaries, a direct combination would produce a false-color image. Instead, we mapped the spectral data to RGB values using CIE color-matching functions, ensuring accurate color representation. After obtaining the correct color, we applied gamma correction to enhance contrast and improve the dynamic range of bright and dark regions. The resulting channels were then combined to reconstruct the final RGB image. For further details on color matching and gamma correction, refer to Supplementary Information, Section III. The results show the ability of our SPI system to work with both coherent and incoherent light sources. To assess the resolution of our system, we used it to image a resolution test target (USAF 1951). The results are shown in Figure 3(c). We imaged the target's central region, resolving group 5, element 3 which demonstrates an excellent resolution of 40.3 lp/mm. This high-resolution capability could enable critical applications in biomedical imaging (e.g. retinal and cellular analysis) [62], industrial quality control for microfabricated components [63], and document authentication where fine-detail discrimination is essential [64].

We next demonstrate the imaging capabilities of the bP-MoS$_2$ photodiode combined with the flexibility of the SPI architecture through an experiment shown in Figure 4(a). A gold-on-glass sample featuring a kangaroo pattern is used as the imaging target, with part of the pattern intentionally obscured by an undoped silicon wafer. Two images are captured: one using a visible LED and the other with the laser operating at a wavelength of 1.55 μm. Despite the opacity of silicon in the visible range, the system successfully reconstructs the obscured portions. This highlights our system capabilities to image through silicon. A black line is observed at the center of the kangaroo in the 1.55 μm image, attributed to light scattering at the edge of the Si wafer.

The second demonstration, shown in Figure 4(b), highlights the edge detection capabilities of our SPI setup. We apply a high-pass filter (HPF) to the sampling voltages recorded by the pre-amplifier to enhance high-frequency components, which correspond to image edges. This simple analog high-pass filtering step enables effective gradient extraction without requiring complex hardware or computational overhead. To convert the gradient map into a binary edge map, we perform magnitude thresholding by applying an adaptive threshold scaled to 70% of the Otsu value [65]. This approach accentuates fine edge details while suppressing the background. The thresholding operation takes approximately 0.55 seconds for full DMD resolution images. Magnitude thresholding is explained in Section IV of the Supplementary Information. As illustrated in Figure 4(b), increasing the HPF cutoff frequency progressively from 100 Hz to 1kHz sharpens the edges in the reconstructed kangaroo image, with the 1 kHz filter yielding the clearest result. These results demonstrate that the system can perform optical edge detection with tunable sensitivity. While the theoretical basis for this approach is discussed in more detail later, the experimental results here confirm its practicality and effectiveness.

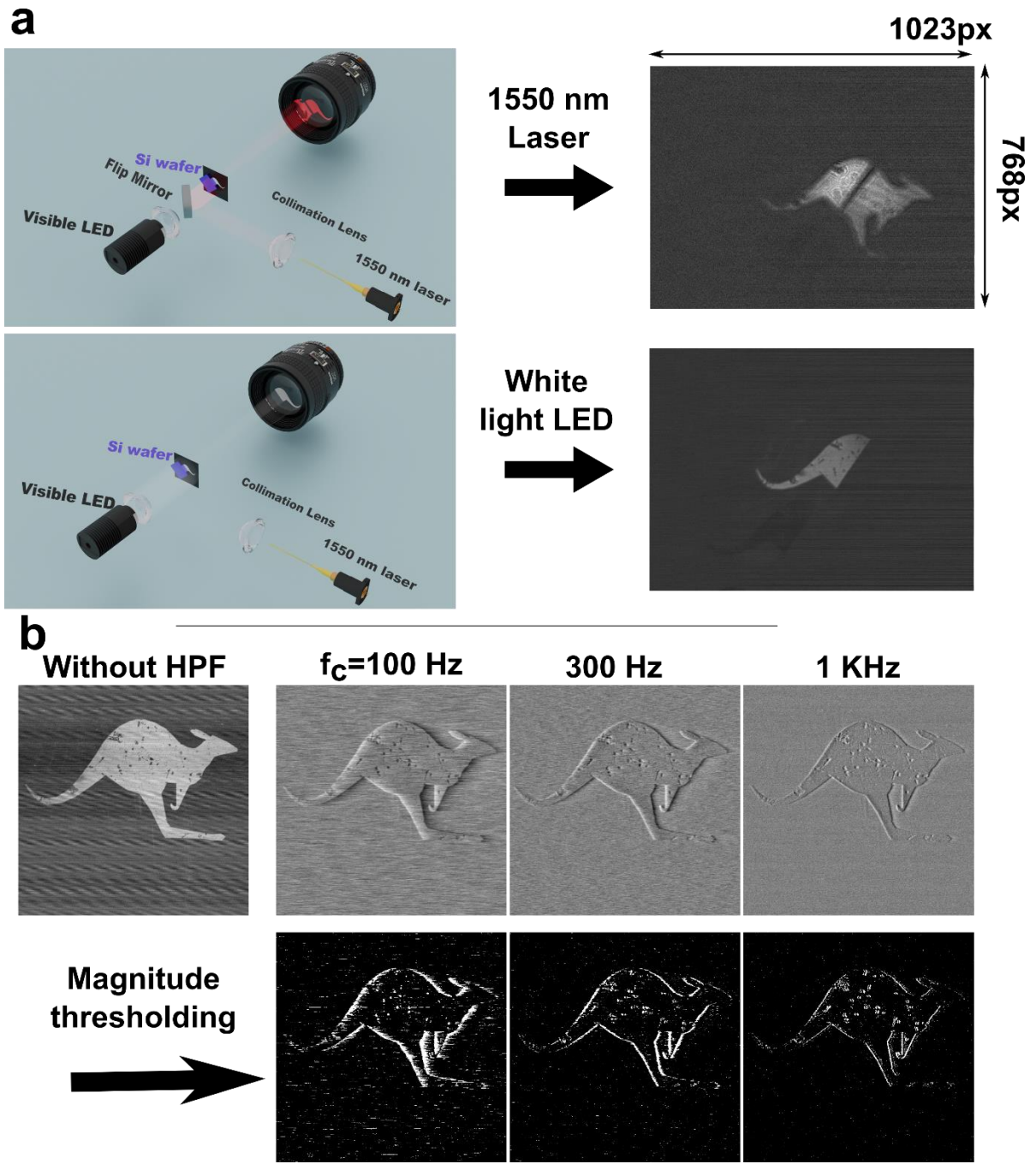

Figure 4. (a) Schematic illustrations and corresponding images from the SPI setup used for imaging through a silicon wafer. Top left: Diagram of the SPI setup with a flip mirror activated to direct 1.55 μm laser illumination onto the silicon wafer, which transmits infrared light. Top right: Image of a kangaroo captured using the SPI system under 1.55 μm laser illumination, demonstrating

successful imaging through silicon. Bottom left: Diagram of the SPI setup with the flip mirror deactivated, allowing white light LED illumination; the visible light is blocked by the silicon wafer. Bottom right: SPI-captured image under white light LED illumination, showing minimal transmission due to silicon opacity in the visible range. (b) Edge enhancement in SPI images using analog HPF of the detected signal. Left: Image captured without HPF, showing a relatively smooth intensity profile. Middle to right top row: Images obtained with HPF applied at cutoff frequencies of 100 Hz, 300 Hz, and 1 kHz, respectively. Bottom row images obtained after magnitude thresholding. It demonstrates increased edge contrast as the cutoff frequency increases. All images in this figure were captured with bP-MoS$_2$ photodiode.

## Discussion:

To evaluate and compare the performance of our vdWs material-based photodetector in the SPI setup with that of a commercial photodetector, we capture a series of images under blue LED illumination using both the bP-MoS$_2$ photodiode and a silicon photodiode. A 100 μm pinhole is placed in front of the commercial Si photodiode to restrict its active area, thereby approximating the detection area of the bP-MoS$_2$ device and ensuring a fair comparison between the two detectors. Imaging is performed across three scenarios: low resolution (127×129 pixels), medium resolution (255×257 pixels), and high resolution (1023×768 pixels). We calculate the PSNR and SSIM of the images captured at different light intensities for both devices. We use the PNG file of the USAF1951 target as the ground truth. The ground truth image and all other images captured are provided in the Supplementary Information section V. Figure 5 (a), (b) shows the PSNR and SSIM comparison between the Si photodiode and the bP-MoS$_2$ photodiode in the SPI setup, respectively. In panel (a) ***, **, and * denote distinct regions of the response curve for the bP-MoS$_2$ device at high resolution. The PSNR and SSIM curves show that, for all scenarios, the PSNR and SSIM increase with increasing light intensities. Lower-resolution images consistently demonstrate higher PSNR and SSIM compared to high-resolution reconstructions under constant light intensity. This behavior can be attributed to two main factors. First, at lower resolutions, more micromirrors are assigned to each image pixel. For instance, in our setup, 36 micromirrors contribute to a single pixel in the low-resolution case in Figure 5 (a), (b). As a result, the total optical signal collected per pixel is higher, leading to a stronger signal relative to the detector noise. Second, the effective number of bits required in the analog to digital conversion process is higher for high-resolution images. Although the total photocurrent and hence the resulting voltage is constant for a given light intensity, higher-resolution images divide this total signal among more pixels, each receiving less optical power. This reduction in per-pixel signal places greater demands on the ADC's precision. In contrast, low-resolution images benefit from a higher signal per pixel, allowing the ADC to operate more effectively within its dynamic range. This second effect is discussed in detail in Supplementary Information VI. Figure 5 (c) presents high-resolution images captured by the bP-$MoS_2$ photodiode under varying light intensities. Stars show corresponding reconstructed images at the intensity levels indicated in (a).

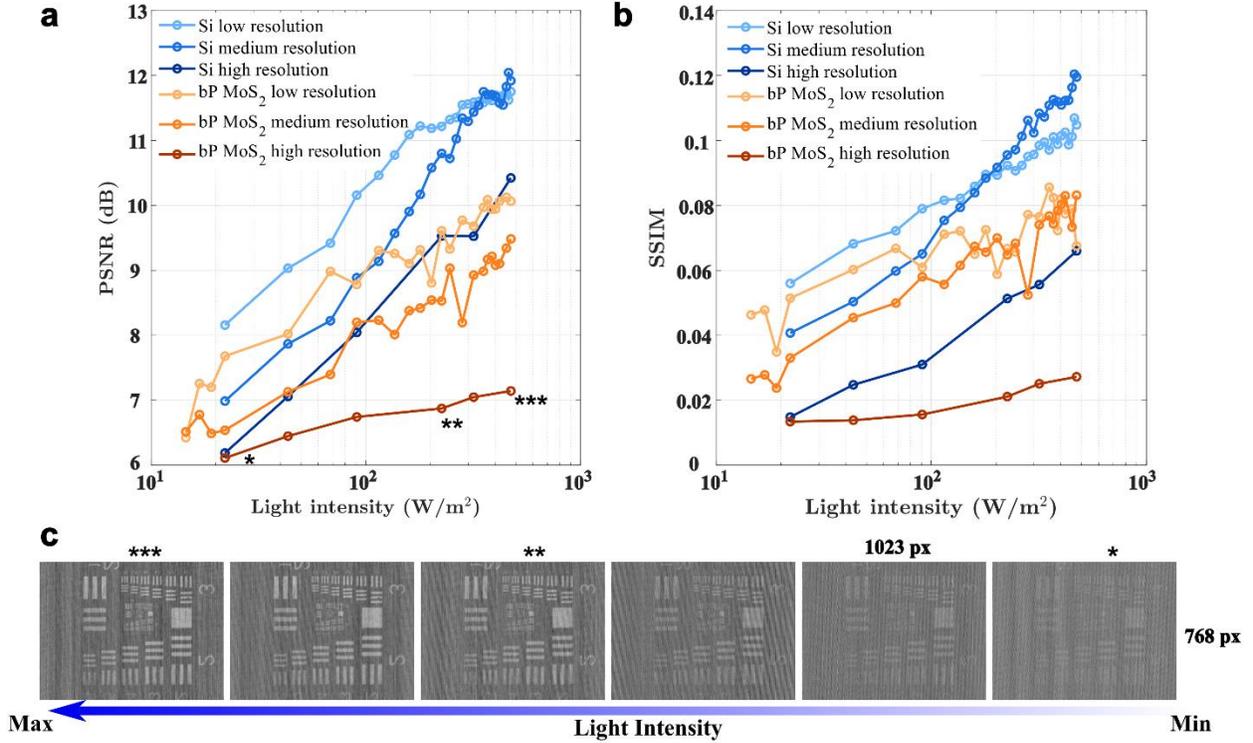

Figure 5. (a) and (b) show PSNR and SSIM of images. These were taken with Si photodiode with 100μm pinhole in front and bP-MoS$_2$ photodiode with SPI setup, respectively. In panel (a) ***, *, and * mark distinct regions of response curve for images taken with the bP-MoS$_2$ device at high resolution. Ground truth image is shown in Supplementary Information section V. (c) High-resolution image captured with bP-MoS$_2$ photodiode at 0.45 μm at different light intensities. Stars show corresponding reconstructed images at intensity levels indicated in (a).

Our sampling strategy employs a cyclic S-matrix derived from the Hadamard matrix [66], constructed using maximal length shift register sequences, producing a first row of order $N = 2^m - 1 = p \times q$, where $N$ is the pixel count, $m > 2$, and $p, q$ define the sampling and image matrix dimensions. The sampling matrix for $(1023 \times 1025)$ is generated by shifting the first row circularly and stored in the random access memory (RAM) of the DMD 7000 board. Section VII of Supplementary Information explains the cyclic S matrix generation and shows its code.

Figure 6 (a) shows an example of a 15 × 17 cyclic S-matrix (N=255). The blue and red dashed rectangles represent the first and second sampling matrices, respectively. These patterns are displayed sequentially on the DMD, with corresponding voltages (e.g., $S_1, V_1$) recorded by the photodiode.

For the reconstruction of the image, we measure voltage samples $V \in \mathbb{R}^N$, which correspond to the circular convolution of the first row of the cyclic matrix $c_1 \in \mathbb{R}^N$, with the imaging vector $X \in \mathbb{R}^N$, scaled by a normalization constant $m$ [36,66]:

$$V = m * (c_1 \circledast X) \quad .(1)$$

When a high pass filter with transfer function $H(f)$ with cut off frequency $f_c$ is applied to the time domain voltage signal, it implicitly high pass filters the image in the pixel frequency domain with cut off frequency $k_c = f_c TN$, in which $T$ is the sampling interval period for each pattern. The filtered image $X_{HP}$ satisfies:

$$DFT\{X_{HP}\}(k) = H_X(k) * DFT\{X\}(k) \quad .(2)$$

Where $DFT$ is the discrete Fourier transform operator and $H_X(k)$ is the high pass filter in pixel frequency domain which combines the effects of $H(f)$ and sampling parameters. Thus, by filtering the measured voltage signal with a high-pass filter, we implicitly apply a frequency-domain high-pass filter to the image $X$, yielding a high-pass filtered version $X_{HP}$. The derivation of the equation (2) and more explanation are in Supplementary Information Section VIII.

Full-resolution single-pixel imaging using the DMD board requires 46 seconds, this results from operating at the DMD at the max frame rate and displaying N frames, where N is the number of pixels. The DMD contains 1024×768 pixels. To capture images with this resolution, we use a 1025×1023 cyclic S-matrix for sampling. This requires 1,048,575 unique patterns to be sequentially displayed. At the DMD's maximum frame rate of 22,727 Hz, this pattern projection process takes approximately 46 seconds, including overheads such as synchronization delays and data acquisition. Reconstruction from the measured data takes less than 50 ms. It is important to note that the edges of the sampling matrix that lie outside the DMD's active area are not projected, resulting in black regions that in our code we remove in the reconstructed image. To reduce acquisition time, we implement a compressed sampling strategy. Instead of displaying all patterns, we skip fixed intervals and interpolated the missing voltages. This is justified by the fact that consecutive sampling patterns differ gradually due to the cyclic structure of the S-matrix. Figure 6(b) schematically illustrates the 50% and 25% sampling schemes, indicating which voltages are directly measured and which are interpolated. Figure 6(c) presents reconstructed images at various sampling rates, along with their corresponding PSNR and SSIM values. Using the 10% sampling case, we achieve a PSNR of 20 dB and an SSIM of 0.179, relative to the 100% sampling image, which serves as the ground truth.

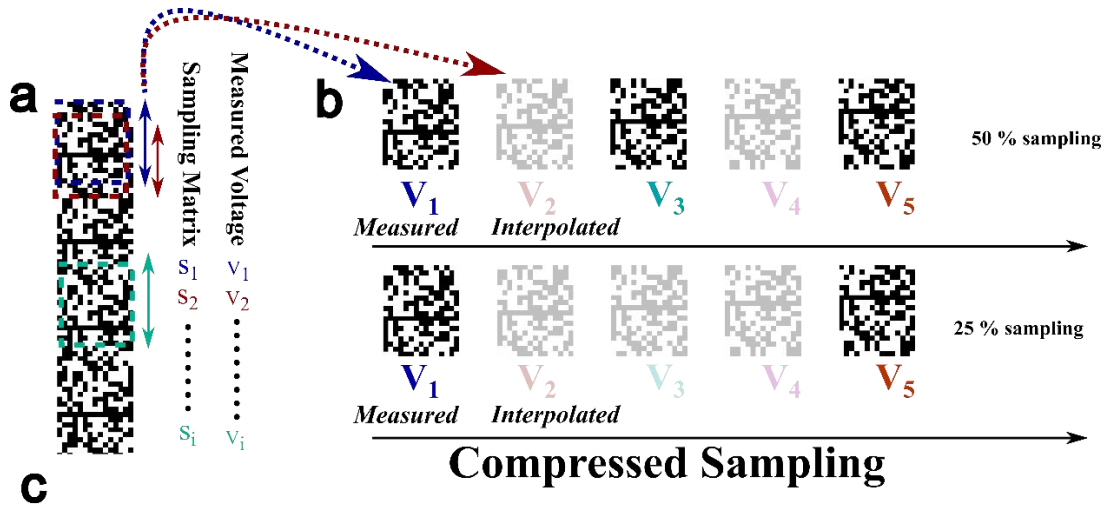

Figure 6. (a) Sampling cyclic S matrix. (b) Proposed compressed sampling technique. (c) Resolution target images captured with bP-MoS$_2$ photodiode under illumination by white LED at different sampling rates. Ground truth image is 100% sampling image.

**Conclusion**:

In this work, we address the limitations of high-resolution imaging from the visible to the ex-SWIR by introducing a scalable and cost-effective SPI system based on a vdWs bP-MoS$_2$ photodiode. Our SPI achieves 1023×768 pixel resolution in the visible and 512×512 pixel resolution in the ex-SWIR, surpassing prior vdWs material-based SPI demonstrations by a factor of 64 in pixel count.

By employing cyclic S-matrix compressed sampling, we reduce imaging time by a factor of four at full DMD resolution. We also introduce a simple yet effective edge detection method for fast feature extraction. In addition, for the first time we demonstrate imaging through silicon wafer using vdWs material-based photodetector SPI. This work demonstrates the viability of vdWs material-based detectors for compact, high-resolution infrared imaging systems, opening opportunities for applications in biomedical diagnostics, gas sensing, autonomous systems, and security technologies.

**Author Contributions**

K. C and SS. M conceived the idea. K.C supervised the research. SS. M and J. C designed the single pixel imaging setup and did the measurement together. S.S.M. wrote the manuscript, except for the color matching and high-pass filter formulation sections in the Supplementary Information, written by J.C.; the fabrication section in the Supplementary Information, written by A.C.; and the focal spot size and its effect on image reconstruction section in the Supplementary Information, written by both S.S.M. and J.C. SS. M did the simulations. S. B, H. L, and S. W helped SS. M and J. C with characterization of the device. A. C, under the supervision of J. B, made the bP-$MoS_2$ photodiode. All authors discussed the results and commented on the manuscript.

Acknowledgement

This work was supported in part by the Australian Research Council (ARC) Centre of Excellence for Transformative Meta-Optical Systems (Project No. CE200100010) and by the ARC Discovery Projects Program (DP240101309). SS.M. acknowledges a scholarship from the University of Melbourne – Shanghai Jiaotong University (SJTU) Joint PhD program.

We thank Luke Philpott for their invaluable assistance during the measurement process. We also thank Nima Sefidmooye Azar, Shaban Sulejman, Mohammad Salehi, and Haiwei Wang for their invaluable support of SS. M in the initial stages of this project.

Supplementary Material for

# High Pixel Resolution Visible to Extended Shortwave Infrared Single Pixel Imaging with a black Phosphorus-Molybdenum di-sulfide (bP-MoS$_2$) photodiode

*Seyed Saleh Mousavi Khaleghi[1,2,3]\*, Jinyuan Chen[1]\*, Sivacarendran Balendhran[1], Alexander Corletto[1], Shifan Wang[1], Huan Liu[2,3,4], James Bullock[1], and Kenneth B. Crozier[1,3,4]©*

[1]Department of Electrical and Electronic Engineering, University of Melbourne, Victoria 3010, Australia

[2]University of Michigan-Shanghai Jiao Tong University Joint Institute, Shanghai Jiao Tong University, Shanghai 200240, China

[3]ARC Centre of Excellence for Transformative Meta-Optical Systems (TMOS), University of Melbourne, Victoria 3010, Australia

[4]School of Physics, University of Melbourne, Victoria 3010, Australia

\* Equal contribution

✉ e-mail: kenneth.crozier@unimelb.edu.au

**Section I. Focal spot size and its effect on image reconstruction:**

We simulate our single-pixel imaging setup to estimate the area illuminated at the Fourier plane of the detection module after the 50× objective lens. Since simulating all mirrors in the DMD is computationally impractical for us, we approximate the DMD's active area by assuming that one row of 1024 mirrors is turned "ON" in the x-direction and one column of 768 mirrors is turned "ON" in the y-direction, forming a cross pattern. The object in our simulation is a square aperture of dimensions 5 mm × 5 mm, chosen such that its image (when projected by the preceding optics) covers a square area of the DMD surface. Each camera lens has an effective focal length of 5 cm, The effective focal length of the 50× objective is 4 mm. Using COMSOL's Ray Optics module for 3D simulation, we find that for the zero-diffraction order of the DMD, at the Fourier plane the rays converge to a root mean square radius of 286 μm after the 50× objective lens. Based on this, we estimate the focal plane spot after 50× objective at the detector plane to be approximately 286 μm in radius.

Fig. S 1(a) presents ray optics simulation results for the complete optical system. In Fig. S 1(a), we show the full ray-tracing layout, while Fig. S 1(b) illustrates cones of rays emanating from various points across the object plane and propagating through the system toward the focal region. Each cone consists of 20 rays in different directions. The opening angle of the ray cone from the object was taken to be 7°, as determined by the numerical aperture (NA = 0.42) of the 50× objective lens. To aid visualization, rays originating from each object point are assigned to a unique color, which remains consistent across all propagation angles. This color-coding allows us to distinguish between the image and Fourier planes. At the image plane, it can be seen that rays from a single point of the object (i.e. coded with the same color) converge to a single point, as expected. At the Fourier plane, it can be seen that rays from different points on the object (i.e. coded with different colors) converge to single points, again as expected.

These rays (converging to a single point in the Fourier plane) emanated at the same angle in the object plane.

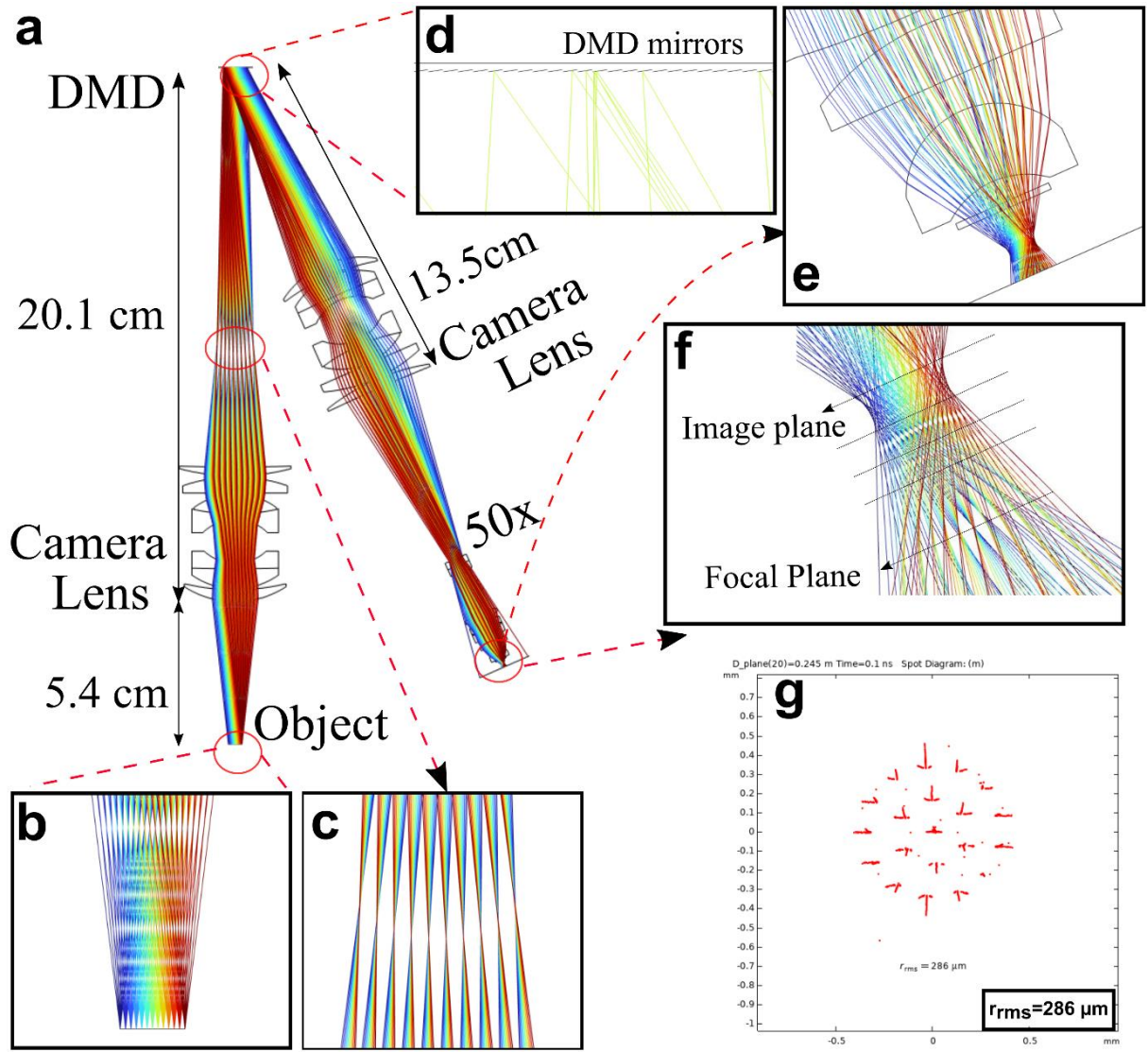

Fig. S 1 (a) Ray optic simulation of our SPI setup. (b) Zoomed in image of illuminated cone of rays from object. (c) Rays at Fourier plane of camera lens. (d) Reflection of rays from DMD mirrors. (e) Zoomed in image of converging rays inside 50× objective. (f) Rays after 50×

objective lens. Dashed lines show the positions of image plane and focal plane. (g) Fourier plane spot diagram.

By examining Fig. S 1(c), it can be seen that this position represents the Fourier plane of the camera lens in the imaging module of the SPI system. Fig. S 1(d) shows ray reflection from the DMD micromirrors, and Fig. S 1(e) shows rays passing through the microscope objective lens. It can be seen that the rays are clipped at the periphery of the 50× objective. This indicates that the 50× objective lens probably functions as the aperture stop of the system. In the detection module of the SPI setup, the distance between the camera lens and the 50× objective was optimized to minimize the spot size at the focal plane of the objective. This optimization was carried out by positioning the Si detector, which has a 100 μm pinhole in front, at the working distance of the 50× objective, and then translating the objective (along with the detector) until the output voltage signal reached a maximum, indicating optimal alignment.

A magnified view of the ray distribution immediately after the 50× objective is shown in Fig. S 1(f). In this region, the image plane appears first, indicated by same-color rays converging, followed by the Fourier plane, where rays of different colors intersect. The beam narrows to its smallest cross-section between these two planes. Fig. S 1(g) shows the spot diagram of the Fourier plane where root-mean-square radius of the spot distribution is calculated to be approximately 286 μm.

Fig. S 2 illustrates the evolution of the beam profile at various axial positions after the 50× objective lens, visualized via spot diagrams. At the image plane Fig. S 2 (a), the spot pattern forms a cross, which replicates the shape of the active micromirror region on the DMD. As the beam propagates forward Fig. S 2 (b-e), the spot distribution narrows and reaches its minimum size at Fig. S 2 (e). Further along the optical axis (Fig. S 2 (f-h)), the rays diverge and eventually

form 20 discrete focal spots at the Fourier plane (Fig. S 2 (h)), where rays from same angle in the object space converge.

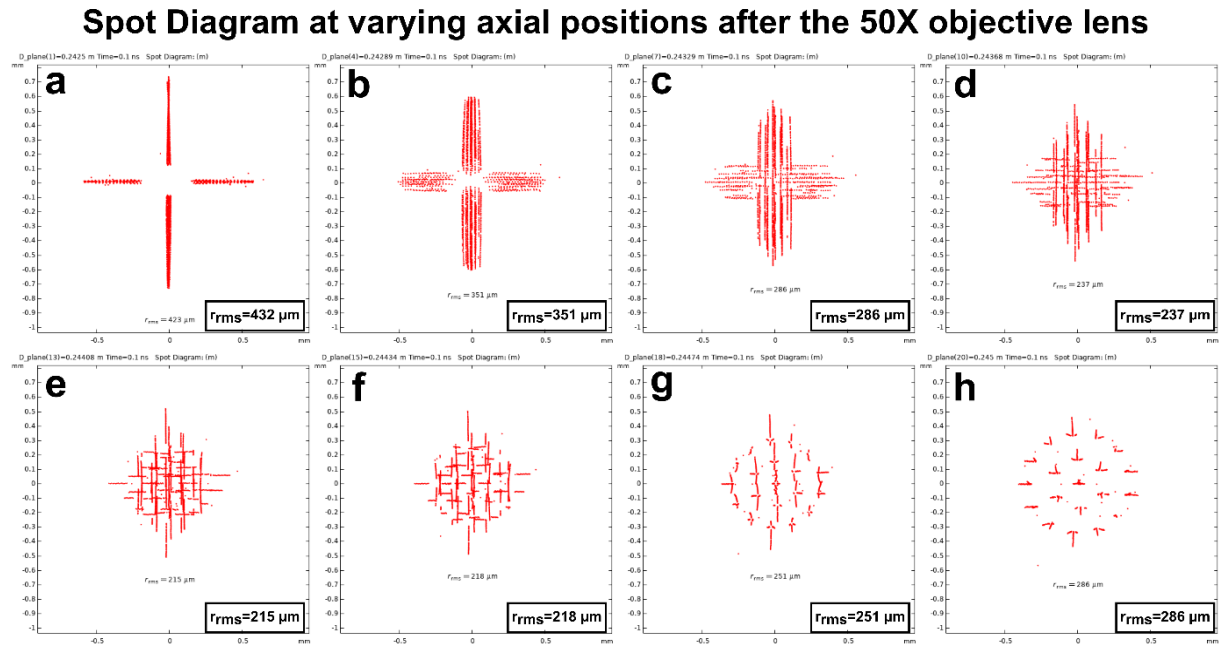

Fig. S 2 Spot diagrams at different axial positions along the optical axis after the 50× objective lens. (a) At the image plane, the spot pattern forms a cross, corresponding to the active cross-shaped micromirror pattern on the DMD surface. (b–e) As the beam propagates, the spot distribution contracts, reaching a minimum at (e). (f–h) Beyond this plane, the beam expands and transforms into a discrete set of focused spots at (h), corresponding to the Fourier plane. Here, parallel rays from different object directions converge into 20 distinct focal spots.

To evaluate how the size of the photodetector's active area affects image resolution in SPI, we performed a simulation using the central region of a standard 2-inch USAF-1951 resolution test chart as the object (Fig. S 3a). This has an extent of 4.8 mm × 4.8 mm. In Fig. S 3b, we plot the Fourier transform log ($|FFT|$), where we use the logarithmic magnitude to enhance dynamic range. To estimate the image size at the Fourier plane, we model the 50× objective lens as a simple lens with a focal length of 4 mm. An object $f(x, y)$ placed at a distance $d$ in front of a lens, under the Fresnel approximation and paraxial conditions, undergoes a Fourier transform. Consequently, at the front focal plane (one focal length away), the field is represented by $F(v_x, v_y)$, the spatial frequency distribution of the object [1]. The intensity distribution $I(x, y)$ is proportional to the squared magnitude evaluated at spatial frequencies $v_x = \frac{x}{\lambda f}$, $v_y = \frac{y}{\lambda f}$ and can be expressed as:

$$I(x,y) = \frac{1}{(\lambda f)^2} \left| F\left(\frac{x}{\lambda f}, \frac{y}{\lambda f}\right) \right|^2 \quad .(1)$$

Where $\lambda$ is the illumination wavelength, and $f$ is the focal length of the lens. The spatial coordinates $(x, y)$ in the Fourier plane are linearly related to spatial frequency $(v_x, v_y)$ in the object by $x = \lambda f v_x$ and $y = \lambda f v_y$

We conducted experiments and determined that the image length before the 50× objective is approximately 4.8 mm. Using the object shown in Fig. S3a, with a resolution of 2048 × 2048 pixels, we obtain a Fourier-plane image (Fig. S3b) with a length of approximately 907 μm, as calculated from eq (1). Several factors restrict the accessible spatial frequencies in the Fourier plane. First, a limit is set by the numerical aperture (NA) of the optical system. Ray-optics simulations (Figs. S1 and S2) show that the illuminated focal-plane region has a diameter of approximately 572 μm, which is indicated by the white circles in Fig. S3b. Second, the pixelated structure of the DMD limits the spatial frequencies that can be resolved by the SPI system. In our configuration, the resolution target is sampled into 768 × 768 pixels, resulting in a Fourier-plane image with a side length of approximately 340 μm, as denoted by the turquoise squares in Fig. S3b. Finally, the finite size of the detector further limits the spatial frequencies that can be captured, as shown by the red circles in Fig. S3b.

Our single pixel detector captures only the central portion of the Fourier plane because our bP-MoS$_2$ photodiode has a spatial extent of approximately 170 μm × 170 μm. This yields satisfactory image reconstruction. When the detector dimensions are much smaller than the focal spot (e.g., <30 μm × 30 μm, typical for many 2D material photodetectors), we observe significant loss of high-frequency details and anisotropic blurring in the reconstructed images. For small detector with size of 30 μm shown in Fig. S 3(c) right panel, the minimum distinguishable lines corresponds to group 2, element 6 of resolution target shown with dash line in Fig. S 3(d) right panel. This resolution corresponds to $7\frac{lines}{mm}$. This severe spatial bandwidth restriction results in heavily blurred reconstruction shown in Fig. S 3(d) right panel.

Conversely, for large detector with size of 170 μm shown in Fig. S 3(c) right panel, the minimum distinguishable bars are from group 5, element 3 of resolution target shown with dash line in Fig. S 3(d) left panel. This resolution corresponds to $40\frac{lines}{mm}$. This enables the system to resolve fine very high frequency details, shown in Fig. S 3(d) left panel.

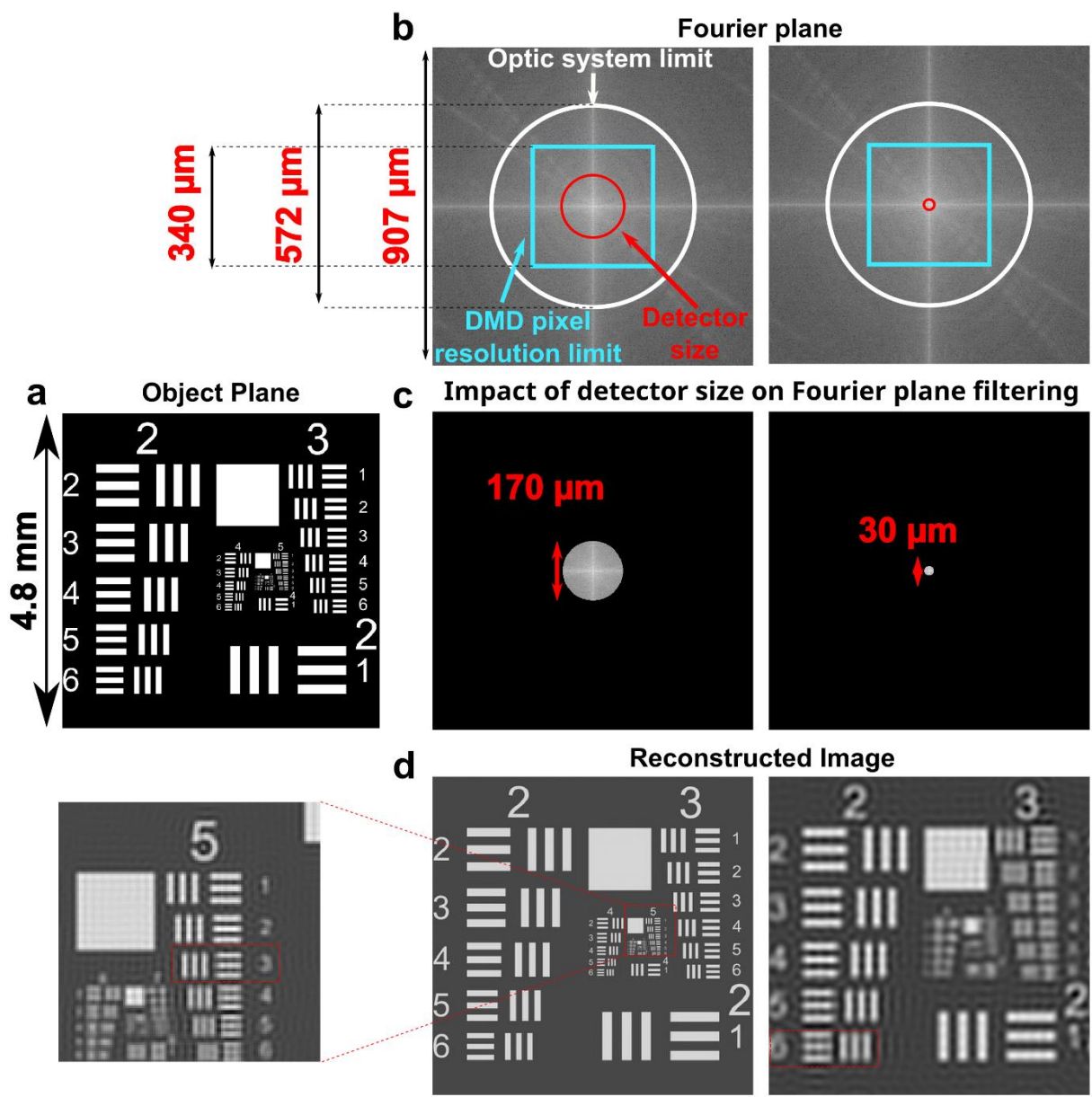

Fig. S 3 Effect of photodetector active area on image reconstruction. (a) Original object image. (b) Fourier plane after f = 4mm lens, with red circles indicating two photodetector active area sizes of 170μm (left) and 30μm (right). (c) FT after spatial filtering with photodetector. (d) Reconstructed images from the filtered FT for both detector sizes, showing degradation with smaller active areas. The red dashed rectangle indicates the minimum resolvable line pairs in each case.

This simulation assumes homogenous photo-absorption over the entire active area, however most of the reports for bP-MoS$_2$ photodiodes [2] show that is not necessarily the case. Although this non-uniform absorption can further impact the reconstructed image quality, a detailed investigation of this effect is beyond the scope of this paper.

**Section II. Details of fabrication process:**

**bP Exfoliation**: We cleaned the Si/SiO$_2$ substrates by bath sonication in acetone and IPA for 10 minutes each, followed by oxygen plasma treatment for another 10 minutes. We gently exfoliated large bP flakes from a bulk bP crystal using thermal release tape. Immediately after exfoliation and plasma cleaning, we pressed the tape with bP flakes onto the Si/SiO$_2$ substrate for 1 minute to ensure good contact. We then heated the substrate to 100 °C to release the tape and leave large bP flakes on the Si/SiO$_2$ substrate.

**MoS$_2$ Exfoliation**: Using a metal-assisted exfoliation method, we exfoliated large-area MoS$_2$ flakes from a bulk MoS$_2$ crystal with white cleanroom tape. The flakes retained a metallic appearance. We deposited a 100 nm Au film on the exfoliated MoS$_2$ on tape via electron beam evaporation starting at 0.1 Å/s for the first 5 nm to minimize damage, then continuing at 1 Å/s for the remainder. We pressed the thermal release tape onto the Au film, ensuring full contact, and then peeled it off, allowing the Au film to preferentially exfoliate mono- or few-layer MoS$_2$ from the thicker flakes. We immediately pressed the thermal release tape carrying the MoS$_2$ flake/Au film onto a cleaned Si/SiO$_2$ substrate for 1 minute. Then, we heated the substrate to 100 °C to release the tape, leaving large MoS$_2$ flakes with the Au film on the Si/SiO$_2$ substrate. Finally, we immersed the substrate in a KI/I$_2$ solution to etch away the Au film, leaving behind large thin MoS$_2$ flakes.

**Transfer of MoS$_2$**: We placed two drops of polycarbonate (PC)/chloroform solution onto a glass slide and pressed another slide on top to form a thin film between them. We stretched this PC film over a 5×5 mm Gelpack PDMS square mounted on a glass slide. Using a translation stage and microscope, we carefully aligned and pressed the PC/PDMS film onto the target MoS flake on the Si/SiO$_2$ substrate, confirmed by a visible color change due to thin-film interference. We then heated the substrate to 100 °C and cooled it to room temperature to ensure conformal contact. We slowly lifted the PC film/PDMS/glass assembly to pick up the MoS$_2$ flake, then transferred it onto the large bP flake substrate. Using the microscope and translation stage, we aligned and pressed the MoS$_2$ flake onto the bP flake. We heated the substrate to 160 °C while lifting the glass slide/PDMS away during heating, leaving the MoS$_2$ flake and PC film on top of the bP flake. To remove residual PC film, we cleaned the bP/MoS$_2$ heterostructure stack in dichloromethane for 30 minutes.

**Contact deposition**: We deposited 60 nm thick Ni contacts on both the bP and MoS$_2$ flakes using electron beam evaporation through a steel shadow mask at a 1 Å/s rate.

**Section III. Color matching and gamma correction:**

In color imaging, we used three LEDs, with wavelengths centered at 780 nm, 565 nm, and 450 nm to illuminate the object in a sequential fashion, corresponding to red, green, and blue spectra. This allowed us to capture three images in these spectral ranges. To reconstruct true color as perceived by the human eye, we applied CIE color mapping techniques. First, we converted the LED colors to XYZ parameters in the CIE color space $(X_{\bar{\lambda}}, Y_{\bar{\lambda}}, Z_{\bar{\lambda}})$ using the 1931 XYZ Color Matching Functions

$$X_{\bar{\lambda}} = \int P(\lambda)\bar{x}(\lambda)\, d\lambda \quad .(2)$$

$$Y_{\bar{\lambda}} = \int P(\lambda)\bar{y}(\lambda)\, d\lambda \quad .(3)$$

$$Z_{\bar{\lambda}} = \int P(\lambda)\bar{z}(\lambda)\, d\lambda \quad .(4)$$

where $P(\lambda)$ represents the LED spectrum and $\bar{x}(\lambda), \bar{y}(\lambda), \bar{z}(\lambda)$ are the CIE XYZ standard observer color matching functions. These $X_{\bar{\lambda}}, Y_{\bar{\lambda}}, Z_{\bar{\lambda}}$ were then transformed into the CIE RGB color space $(R_{\bar{\lambda}}, G_{\bar{\lambda}}, B_{\bar{\lambda}})$ through a linear transformation matrix M, allowing each LED spectrum to be expressed as a linear combination of R, G, and B coordinates. Hence the r, g, and b value for each pixel of the image can be represented as:

$$[r\ g\ b] = [1\ 1\ 1] \times [a\ b\ c]^T \times [RGB_a\ RGB_c\ RGB_c] \quad .(5)$$

Where a, b and c are measured intensities at $\lambda$=780 nm, 565 nm and 450 nm respectively, $RGB_a$, $RGB_b$, $RGB_c$ are row vectors, each containing the RGB color information corresponding to its respective wavelength.

Finally, the measurements from the three-color LEDs from each pixel were converted to RGB values and gamma correction was applied to adjust the brightness and contrast for human visual perception, resulting in a reconstructed color image.

**Section IV. Magnitude Thresholding:**

To extract binary edge maps from the gradient images obtained after high-pass filtering, we employed a magnitude thresholding procedure. First, the absolute value of the gradient image was computed to obtain the gradient magnitude, discarding directional information. This magnitude map was then normalized to the [0, 1] range by dividing all pixel values by the global maximum. This normalization ensures consistent thresholding performance across different images or filter settings.

To convert the normalized gradient magnitudes into a binary edge map, we applied an adaptive thresholding technique based on Otsu's method [3]. Specifically, we used MATLAB's "graythresh" function to determine the optimal global threshold for separating foreground (edges) from background. To enhance sensitivity and capture weaker edges, the threshold was scaled to 70% of the Otsu value before binarization using the "imbinarize" function. This procedure was applied to both the positive and negative polarity of the gradient, effectively inverting the image contrast to capture edges in both bright-to-dark and dark-to-bright transitions. The result is a high-contrast binary image highlighting edge locations. This post-processing step is computationally lightweight, requiring approximately 0.7 seconds for a full-resolution image.

**Section V. Images captured at different intensities at varying pixel resolutions:**

Fig. S 4 shows images captured using a blue light LED with peak emission at a wavelength of $\lambda = 0.45$ μm, employing both a commercial silicon photodiode with a 100 μm diameter pinhole and our bP-MoS$_2$ photodiode, across various light intensities and different pixel resolutions. The reported light intensity refers to the intensity at the detector plane. To evaluate image quality, we calculated the PSNR and SSIM based on the ground truth image.

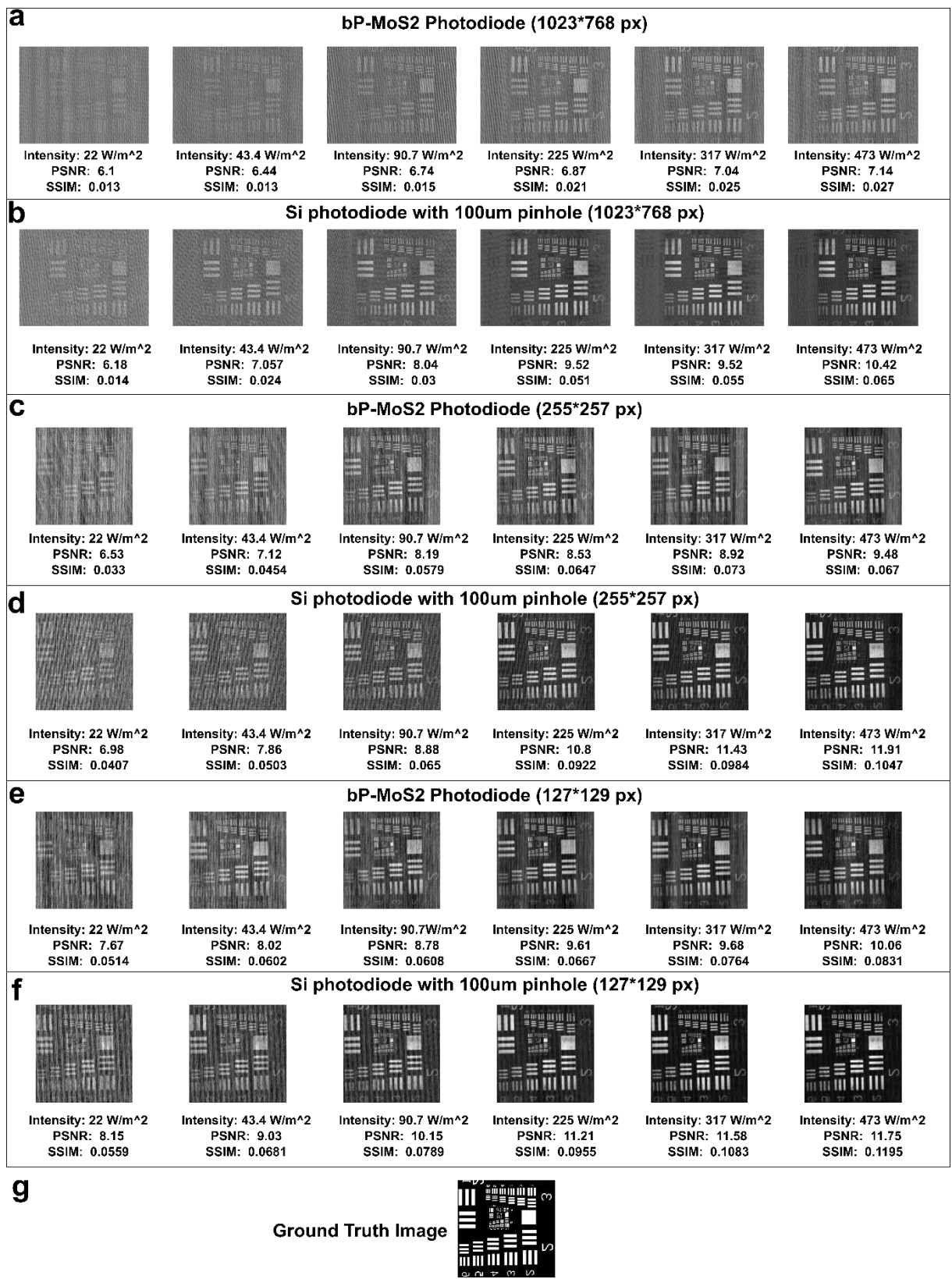

Fig. S 4 (a) Images captured using bP-MoS2 photodiode at different light intensities. (b) images captured by Si photodiode with a 100um pinhole in front.

**Section VII. Effective Number of Bits Required for High-Resolution:**

Here we calculate the effective number of bits required for high resolution single pixel imaging with the cyclic S-matrix approach. For this purpose, we use MATLAB to multiply each sampling pattern to the target image. We use the USAF1951 resolution image as the target image. We then sum up these products to calculate the required number of voltage levels.

The image is represented as a vectorized matrix $X_{1 \times N}$, where $N$ is the total number of pixels. The sampling matrix is denoted as $S_{N \times N}$, and its j$_{th}$ row, $S_{j_{(1 \times N)}}$, corresponds to j$_{th}$ sampling pattern. The photodiode output voltage for the j$_{th}$ measurement is linearly proportional to the inner product (·) between the image vector $X$ and the sampling vector $S_j$:

$$P_j = S_j \cdot X \qquad (6)$$

The number of voltage levels required corresponds to the number of unique values in $P_j$. Fig. S 5 (a) shows the original image, and (b) presents the reconstructed image obtained using the S-matrix approach. For this example, the number of unique values in P is 329633, which corresponds to an effective ADC resolution of $\log_2(329633) \approx 18.33$ bits. In comparison, the lower resolution image requires only 11108 unique levels, corresponding to an ADC resolution of approximately 13.4 bits.

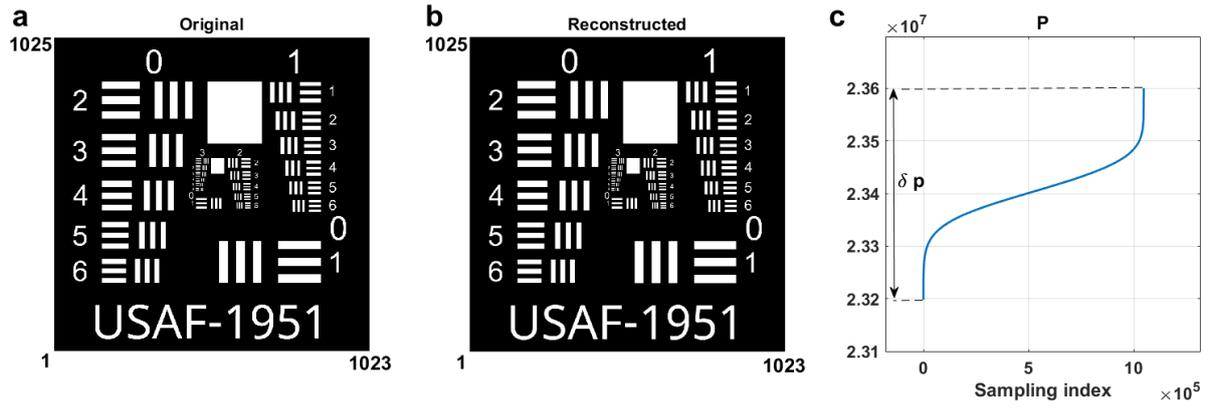

Fig. S 5 a). Original input image. (b) Simulated reconstructed image using cyclic-S-matrix. (c) Inner product of the image vector $X$ and the sampling vector $S_j$ versus sampling index $j$.

**Section VII. Cyclic S Matrix Generation and Code:**

The procedure of producing a cyclic S matrix of the order $N = 2^n - 1$ $(n = 1,2,3,...)$ is as follows:

The first row of S should be generated using a "maximal length shift register sequence" of length $N$. For generating this sequence, we require a binary primitive polynomial of degree $n$ [4].

Table 1 The binary primitive polynomials

| n | Polynomial | n | Polynomial |
|---|---|---|---|
| 9 | $a^9 + a^4 + 1$ | 10 | $a^{10} + a^3 + 1$ |
| 11 | $a^{11} + a^2 + 1$ | 12 | $a^{12} + a^7 + a^4 + a^3 + 1$ |
| 13 | $a^{13} + a^4 + a^3 + a + 1$ | 14 | $a^{14} + a^{12} + a^{11} + a + 1$ |
| 15 | $a^{15} + a + 1$ | 16 | $a^{16} + a^5 + a^3 + a^2 + 1$ |
| 17 | $a^{17} + a^3 + 1$ | 18 | $a^{18} + a^7 + 1$ |
| 19 | $a^{19} + a^6 + a^5 + a + 1$ | 20 | $a^{20} + a^3 + 1$ |

This shift register produces the largest period possible with n stage shift registers. Once the first row of the cyclic matrix **S** is generated, the remainder of the matrix is formed by circularly shifting the row one position to the left at each step. Table 1 shows the binary primitive polynomials of degree 9 to 20, which are used to generate the maximum-length sequences of "0"s and "1"s using $m$ shift registers. Fig. S 6 shows MATLAB code that can be used to produce the first row of the cyclic S-matrix for n=20.

```
%Maximum Length Shift Register Sequences for n

N=2^n-1;

S=[];

s=[zeros(1,n-1) 1]; %initial value of s

Res=zeros(1, N+2*n);

for i=1:floor(N)+1
```

|  %  x^20+x^3 + 1           | %x^19+x^6+x^5+x+1                |
| -------------------------- | -------------------------------- |
| Res(i)=s(1);               | Res(i)=s(1);                     |
| sum0= mod(s(1)+s(4),2);    | sum0=mod(s(1)+s(2)+s(6)+s(7),2); |
| s=[s(2:end) sum0];         | s=[s(2:end) sum0];               |

```
end
```

Fig. S 6 MATLAB code used to generate first row for cyclic Smatrix.

The cyclic **S** matrix needs to be tiled to generate patterns suitable for single-pixel imaging. The tiling can be done based on the factors of N, where $N = 2^n − 1$. For example, if $n = 4$, then $N = 15$, and the factors of 15 are 1, 3, 5, and 15. This allows us to arrange the cyclic **S** matrix into tiled patterns such as a 3×5 rectangular array, as illustrated in Fig. S 7. In this figure, each highlighted rectangle (green, yellow, and red) contains 15 pixels and demonstrates a one-pixel cyclic shift from the others. The black pixels represent binary "0" values, and the white pixels represent binary "1" values, corresponding to the structure of the cyclic matrix.

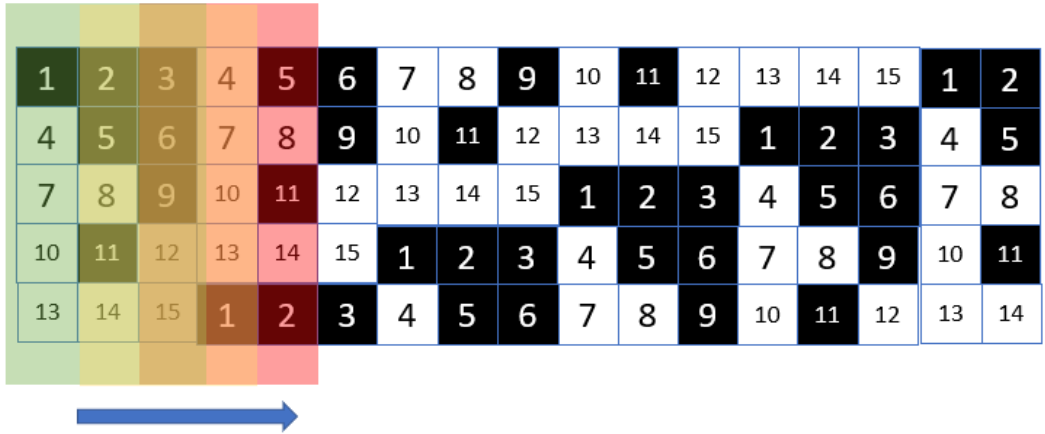

Fig. S 7 Cyclic S matrix for 15 pixels.

## Section VIII. Temporal High-pass Filtering as Spatial Frequency Filtering in Single-Pixel Imaging:

For the reconstruction of the image, we measure voltage samples $V \in \mathbb{R}^N$, which correspond to the circular convolution of the first row of the cyclic matrix $c_1 \in \mathbb{R}^N$, with the imaging vector $X \in \mathbb{R}^N$, scaled by a normalization constant $m$ [4,5]:

$$V = m * (c_1 \circledast X) \qquad .(7)$$

The temporal voltage measurement $v(t)$ is a time-multiplexed measurement of spatial information that can be modelled as a sequence of rectangular pulses, each representing the voltage reading for a single pattern within a fixed-duration window:

$$v(t) = \sum_{i=1}^{N} V(i) * rect\left(\frac{t-\left(i-\frac{1}{2}\right)T}{T}\right) \qquad .(8)$$

The Fourier transform of the time-domain voltage signal $\hat{v}(f)$ is defined as [6]:

$$\hat{v}(f) = F\{v(t)\} = \int_{-\infty}^{\infty} v(t) e^{-j2\pi f t} \, dt \qquad .(9)$$

Where $f$ denotes the temporal frequency in Hz.

Hence, the Fourier transform of the voltage signal is given by:

$$\hat{v}(f) = F\{v(t)\} = \sum_{i=1}^{N} V(i) * T * sinc(fT) * e^{-j2\pi f\left(i-\frac{1}{2}\right)T} \qquad .(10)$$

This expression can be simplified as:

$$\hat{v}(f) = T * sinc(fT) * e^{j\pi fT} \sum_{i=1}^{N} V(i) * e^{-j2\pi fTi} \qquad .(11)$$

Each voltage sample $V(i)$ corresponds to a structured projection of the object onto a shifted pattern from the cyclic matrix, as shown in Eq. (8). For the sequence of voltage samples $V(i)$ in the spatial domain, we use the discrete Fourier transform (DFT) defined as [6]:

$$DFT\{V\}(k) = \sum_{i=1}^{N} V(i) * e^{-j2\pi ki/N} \qquad .(12)$$

Therefore, the Fourier transform of the time-domain voltages signal, as shown in Eq. 11))contains the DFT of the voltage samples $V(i)$, evaluated at the spatial frequency integer $k = fTN$, such that:

$$\hat{v}(f) = T * sinc(fT) * e^{j\pi fT} * DFT\{V\}(fTN) \qquad .(13)$$

When a high-pass filter with transfer function $H(f)$ is applied to the time domain voltage signal, the resulting Fourier transform is:

$$\hat{v}_{HP}(f) = H(f) * \hat{v}(f) = T * sinc(fT) * e^{j\pi fT} * H(f) * DFT\{V\}(fTN) \qquad .(14)$$

Substituting $\boldsymbol{V} = m * (\boldsymbol{c_1} \circledast \boldsymbol{X})$ we obtain:

$$\hat{v}_{HP}(f) = T * sinc(fT) * e^{j\pi fT} * H(f) * DFT\{m * (\boldsymbol{c_1} \circledast \boldsymbol{X})\}(fTN) \qquad .(15)$$

Using the DFT convolution property, this becomes:

$$\hat{v}_{HP}(f) = T * sinc(fT) * e^{j\pi fT} * m * DFT\{\boldsymbol{c_1}\}(fTN) * H(f) * DFT\{\boldsymbol{X}\}(fTN) \qquad .(16)$$

Let us now consider the application of a spatial frequency domain high pass filter to the image to produce a high-pass filtered image $\boldsymbol{X_{HP}}$. Its DFT can be expressed as:

$$DFT\{\boldsymbol{X_{HP}}\}(k) = \sum_{i=1}^{N} X_{HP}(i) * e^{-j2\pi ki/N} \qquad .(17)$$

The $DFT$ of the high-pass spatial filtered image $X_{HP}$ can be written as:

$$DFT\{\boldsymbol{X_{HP}}\}(k) = H_X(k) * DFT\{\boldsymbol{X}\}(k) \qquad .(18)$$

Consider the case that the spatial frequency domain high pass filter is related to the temporal domain high pass filter as follows:

$$H_X(k) = H\left(\frac{k}{TN}\right) \qquad .(19)$$

In this case, Eq. (11) can be rewritten as:

$$\hat{v}_{HP}(f) = T * sinc(fT) * e^{j\pi fT} * m * DFT\{\boldsymbol{c_1}\}(fTN) * DFT\{\boldsymbol{X_{HP}}\}(fTN) \qquad .(20)$$

This demonstrates that applying a high-pass filter $H(f)$ in the temporal domain to the measured signal $v(t)$ is mathematically equivalent to applying a spatial frequency domain filter $H_X(k)$ to the image $X$ where the filters are related by Eq. (19).